# 关于金银复合纳米晶的合成顺序与合成机理的实验研究

吉林大学物理学院 白子豪[a] 王一博[b] 刘庆辉[c]

## 摘要：

在本实验中，作者对金银纳米晶合成与表征的原理及方法进行了综述，提出了两种可能的金银复合纳米晶合成机制。作者采用"柠檬酸钠无种子法"（Turkevich Method）对单质金纳米晶、单质银纳米晶、金银复合纳米晶进行制备，并采用动态光散射(DLS)、紫外可见光光谱（UV-Vis）、扫描电子显微镜（SEM）对产物进行表征。通过对比实验，作者对金银复合纳米晶不同合成顺序背后的反应机理问题展开探讨。并对实验中相关新奇现象展开分析。

## 背景与原理：

纳米材料由于其独特的表面效应与量子尺寸效应，在包含力学、热学、电学、磁学、光学、催化等诸多领域具有重要的应用价值。因此，对不同尺寸、不同形貌、不同组分、不同结构的纳米晶的合成与表征方法的研发与优化具有着极高的重要性。对于金、银纳米晶以及其复合物而言，人们已有数十年的研究历史，成功的合成出各种形貌、尺寸的结构，如球形、棒状、立方体、凹形立方体、二十四面体、类球形核壳结构、立方体核壳结构等；在合成方法上主要分成有种子生长法、无种子生长法。[1]在表征方法上，对于电子行为主要采取紫外可见光光谱（UV-Vis）对纳米晶的等离子共振吸收峰进行测量，这是由于不同形貌、尺寸的纳米晶由于各原子电子之间的相互作用，往往使纳米晶表现出不同的等离子共振行为；对于纳米晶的形貌，主要采取扫描电子显微镜（SEM）、透射电子显微镜（TEM）、高分辨透射电子显微镜对形貌进行直接表征。

Turkevich 等人在 1951 年采用柠檬酸钠还原的方式制备纳米金，被称为 Turkevich Method[2]。这是一种无种子生长法。无种子生长法制备流程简单、原理清晰，采用柠檬酸钠、硼氢化钠等还原剂对氯金酸、硝酸银等反应前体进行还原成核，并通过控制反应时间、反应环境，从而控制晶体的生长，进而控制产物的形貌与尺寸。其中柠檬酸钠等药品还起到与纳米晶配位、进而依附在晶体表面，使生长过程稳定可控、防止二次成核等作用。在无种子金纳米晶合成过程中，常常采用硝酸银作为催化剂。然而由于"Galvanic exchange"效应，溶液中的三价 Au 离子可以与 Ag 单质发生氧化还原反应。而 Au 和 Ag 都可以被柠檬酸钠所还原。故以不同的次序、浓度添加硝酸银，对反应产物的形态与性质具有显著影响。例如，通过 Au 离子对 Ag 原子的氧化，可以制备"银核金壳"的纳米结构，以及"金银纳米笼"结构[3]。

a, b 为共同第一作者；c 为通讯作者，通讯作者邮箱为 liuqinghui@jlu.edu.cn

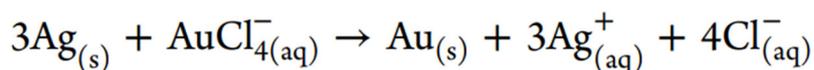

$$3Ag_{(s)} + AuCl_4^-{}_{(aq)} \rightarrow Au_{(s)} + 3Ag^+{}_{(aq)} + 4Cl^-{}_{(aq)}$$

反应 1：Galvanic exchange 效应。溶液中的三价 Au 离子可以与 Ag 单质发生氧化还原反应[3]。

然而，纳米金、银的添加顺序、添加浓度对产物的影响以及其背后的反应机理一问题，具有更广阔的探索空间。在不同合成顺序下，所控制合成的丰富多彩的纳米结构亦具有重要的科研价值与应用价值。在本实验中，我们对"银离子对金纳米晶合成的影响"提出两个机理假设：1，银离子被柠檬酸钠还原后黏附在已经形成的金纳米晶上；2，银离子被柠檬酸钠还原后形成新的银纳米晶，其表面可能进一步与金离子反应，在表面形成金原子壳层。而后我们通过设计不同合成顺序的实验，采取 SEM、UV-Vis、动态光散射（DLS）进行表征，对这两个机制进行探究。

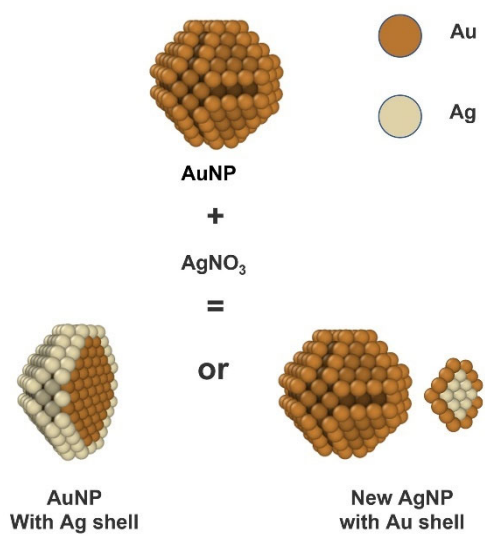

图 1: 本文提出的两个反应机理假设。左图：银离子被还原后黏附在现有的金纳米晶上；右图：银离子被还原后形成新的银纳米晶，并可能进一步被金离子在表面还原，形成金原子壳层。

## 方法:

我们共进行了 8 组实验，每组实验中，采用氯金酸作为金的前体，采用硝酸银作为催化剂以及银的反应前体，以柠檬酸钠作为还原剂。当柠檬酸钠与反应前体在沸水中时，其发生氧化还原反应，成核及生长过程开始快速进行；当反应环境冷却至室温时，反应停止或缓慢进行；同时，氯金酸还可以将单质银氧化，从而对银纳米晶进行置换；因此我们通过控制反应的温度与时间从而控制晶体生长的尺寸、通过控制反应前体的浓度以及金银前体加入的顺序控制纳米晶的结构。

我们对反应后的溶液取样进行动态光散射表征，动态光散射通过颗粒对激光光子的散射数据，对一定温度、粘度的介质下粒子的布朗运动情况进行拟合，通过多普勒效应得到溶液颗粒的尺度大小。在动态光散射过程中，我们充分的进行了溶液的分散，并控制激光强度，从而控制散射的光子数，避免光子的二次散射。

同时我们对反应后的溶液抽样进行紫外可见光光谱测量，在测量过程中，充分的对溶液

进行了分散。

在进行扫描电子显微镜成像时，我们首先对样品进行离心提纯，并将样品滴在用导电胶粘附在电镜底座上的硅片上，待风干后进行电镜成像。提纯是为了避免溶液中其他组分对样品的导电性的影响，从而提高电镜的清晰度；采取硅片是为了避免电子束轰击过程中，基质升温变形，使成像非静态，同时采取硅片也保证基质足够干净，避免其他物质对成像影响。

具体的实验步骤以及部分现象见表1所示，UV-Vis数据如图1所示，DLS图像见附录1。

| 组数 | 步骤序号 | 步骤或现象 | 1 wt%柠檬酸钠/ml | 0.5 wt%氯酸金/ml | 0.1 wt%硝酸银/ml | 1 wt%硝酸银/ml | 加水后体积/ml |
|---|---|---|---|---|---|---|---|
| 纳米金1 | 0min | 将混合物加入沸水 | 2 | 3 | 0.1 | | 100 |
| | 13min30s | 向体系继续加入药剂 | | | 少量 | | |
| | 21min30s | 取样 | | | | | |
| | DLS | 如DLS图像Gold1：小部分20-30nm，大部分100nm | | | | | |
| | 提纯 | 9000rad/s离心15min，取上清液以1：3的比例加入丙酮，以10000rad/s离心15min，取上清液 | | | | | |
| | DLS | 如DLS图像Gold2：20-30nm | | | | | |
| | UV | 光谱数据"Gold1" | | | | | |
| | SEM | SEM组Gold | | | | | |
| | 室温放置半个月 | | | | | | |
| | DLS | 100nm左右 | | | | | |
| | SEM | SEM组Gold_2, SEM观察粒径在1微米至10微米量级与DLS结论不符 | | | | | |
| 纳米银1 | 0min | 将柠檬酸钠加入沸腾硝酸银 | 1.5 | 0 | | 1 | 47.5 |
| | 0-10min | 无色 | | | | | |
| | 10min | 浅黄 | | | | | |
| | 10-30min | 溶液颜色逐渐加深 | | | | | |
| | 30min | 黄褐色 | | | | | |
| | 1h | 取样 | | | | | |
| | DLS | 如DLS图像Silver：粒径50nm | | | | | |
| | UV | 光谱数据"Silver1" | | | | | |

| 阶段 | 时间 | 操作/现象 | | | | | |
|---|---|---|---|---|---|---|---|
| 金核银壳1 | 0min | 加药剂 | 2 | 0.5 | 0.05 | | 60-70 |
| | 1h | 加药剂 | 1 | | | 1 | 50 |
| | 2h | 取样 | | | | | |
| | 提纯第一步 | 离心机9000rad/s，取上清液加丙酮，发现明显的分层现象，上层深粉红色与纳米金溶液相似，下层黄色与银纳米晶溶液相似 | | | | | |
| | 提纯第三步 | 将固体溶解，9000rad/s分别离心 | | | | | |
| | DLS | 如DLS图像Gold-Silver1：少部分20-30nm，大部分100-300nm | | | | | |
| | UV | 光谱数据"Gold-Silver1" | | | | | |
| | SEM | SEM组Gold-Silver | | | | | |
| 金核银壳2（纳米金提纯后加入银前体） | 加药品，混合5分钟加入沸水 | 立刻变浅粉色，之后颜色加深 | 1.5 | 1 | 0.043 | | 50 |
| | 1min | 紫黑 | | | | | |
| | 2min | 紫红 | | | | | |
| | 5min | 酒红 | | | | | |
| | 1h | 取样 | | | | | |
| | DLS | 5% 20nm 90%100-150 | | | | | |
| | 提纯 | 加入丙酮，离心机8000rad/s 15min 取上清液 | | | | | |
| | DLS | 20-30nm | | | | | |
| | 加药品 | 加入后颜色变化，从浅粉色变成紫色 | 少量 | | | 1 | |
| | 室温放置12h | 取样 | | | | | |
| | DLS | 如DLS图像Gold-Silver2：上千 | | | | | |
| | UV | 光谱数据"Gold-Silver2" | | | | | |
| 银纳米晶2（增加硝酸银浓度）、银核金壳1 | 0min | 加药品，混合后立即变乳白色 | 1.5 | | | 1 | |
| | 0-3min | 加入沸水，刚开始无色 | | | | | 50 |
| | 3min | 颜色逐渐加深为浅黄色 | | | | | |
| | 3-8min | 颜色继续加深 | | | | | |
| | 8min | 颜色变为灰黄色 | | | | | |
| | 10min | 焦黄，灰黑，分层了 | | | | | |
| | 20min | 乳黄色 | | | | | |
| | 50min | 停止加热至冷却，取样 | | | | | |
| | DLS | 如DLS图像Silver：50nm左右 | | | | | |
| | UV | 光谱数据"Silver2" | | | | | |
| | 加药品 | 颜色立即变为蓝黑色 | | 0.5 | | | 50 |
| | 加热搅拌50分钟，取样 | | | | | | |
| | DLS | 如DLS图像Silver-Gold1：50nm左右 | | | | | |
| | UV | 光谱数据"Silver-Gold1"（峰的位置与之前的银纳米晶几乎相同，展宽变大） | | | | | |
| 银核金壳2（减少硝酸银） | 0min之前 | 加药品，加热直至沸腾 | | | | 0.5 | 48 |
| | 0min | 加药品，发现特别呛 | 1.5 | | | | |
| | 3min | 浅黄 | | | | | |
| | 3-5min | 加深 | | | | | |
| | 5min | 黄色 | | | | | |
| | 10min左右 | 黄褐色逐渐浑浊 | | | | | |
| | 1h | 取样 | | | | | |
| | DLS | 50nm左右 | | | | | |
| | 1h | 加药品（没有混合直接加入），瞬间灰黑色 | 1 | 0.5 | | | 补水 |
| | 2h | 但这次颜色偏焦黄而非灰黑 取样 | | | | | |
| | DLS | 如DLS图像Silver-Gold2：50nm左右 | | | | | |
| | UV | 光谱数据"Silver-Gold2" | | | | | |

| 银核金壳3（增加金前体、增加柠檬酸钠，减少反应时间） | 0min | 向沸水中加入药品，加入后继续加热直到沸腾 | 15 | | | 1 | 50 |
| --- | --- | --- | --- | --- | --- | --- | --- |
| | 5min | 黄色到焦黄到灰黄色 | | | | | |
| | 10min | 取样 | | | | | |
| | DLS | 如DLS图像Silver-Gold3_1:50nm | | | | | |
| | 30min | 浑浊灰蓝色 | | | | | |
| | 1h | 蓝灰色，取样 | | | | | |
| | DLS | 如DLS图像Silver-Gold3_2: 百分之40 100nm 百分之60 1000nm | | | | | |
| | 重新计时0min | 取一毫升10min的反应物稀释到60ml左右，加入药品室温重新加热到沸腾，银纳米晶稀释后是黄色的，加入氯金酸后颜色变暗一些的黄色 | | 4 | | | 60 |
| | 5min | 温度升到80°左右，颜色变成灰色，而后泛红 | | | | | |
| | 10min | 变成偏黄的非透明的咖啡色，泥土色，此时已经沸腾 | 0 | | | | |
| | 12min | 取样 | | | | | |
| | DLS | 如DLS图像Silver-Gold3_3:157nm | | | | | |
| | UV | 光谱数据"Silver-Gold3-10min" | | | | | |
| | 50min | 取样 | | | | | |
| | DLS | 如DLS图像Silver-Gold3_4:少部分70nm，大部分1370nm | | | | | |
| | UV | 光谱数据"Silver-Gold3-50min" | | | | | |
| | 50min | 对之前60min的银纳米晶 加入4ml氯金酸加热至沸腾，过程中颜色加深至黑色，而后出现深酒红色 | | | | | |

表1：实验步骤以及部分现象

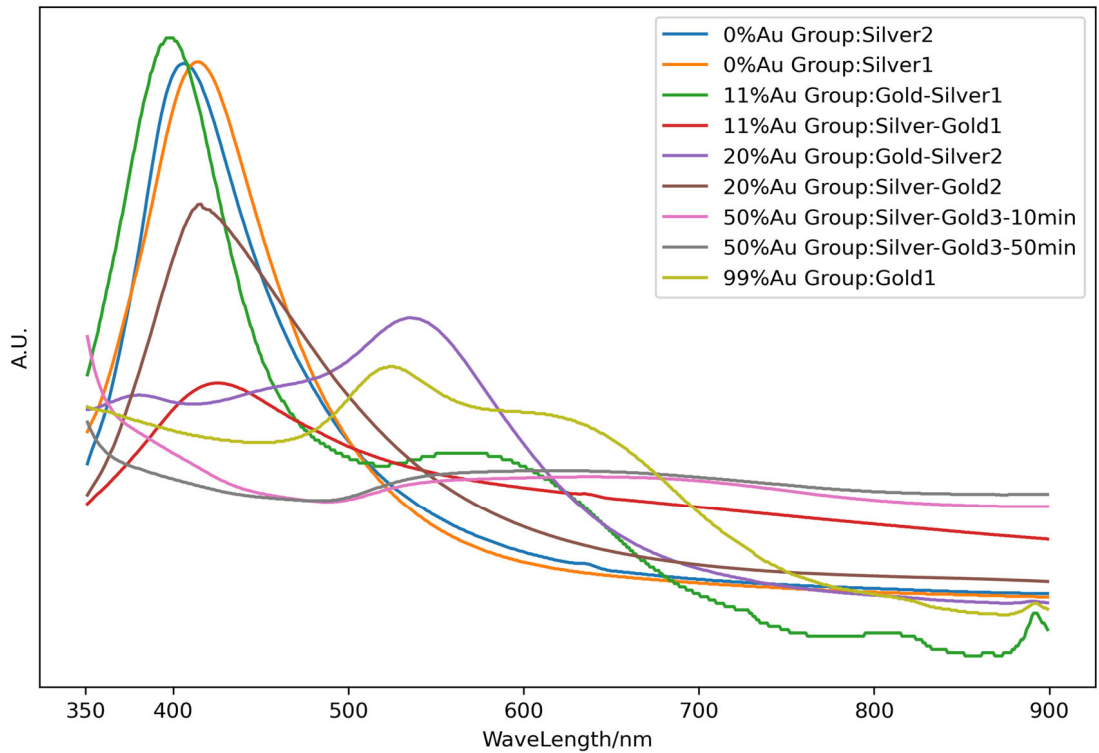

图2：所有实验组的紫外可见光光谱（UV-Vis）数据（已归一化）

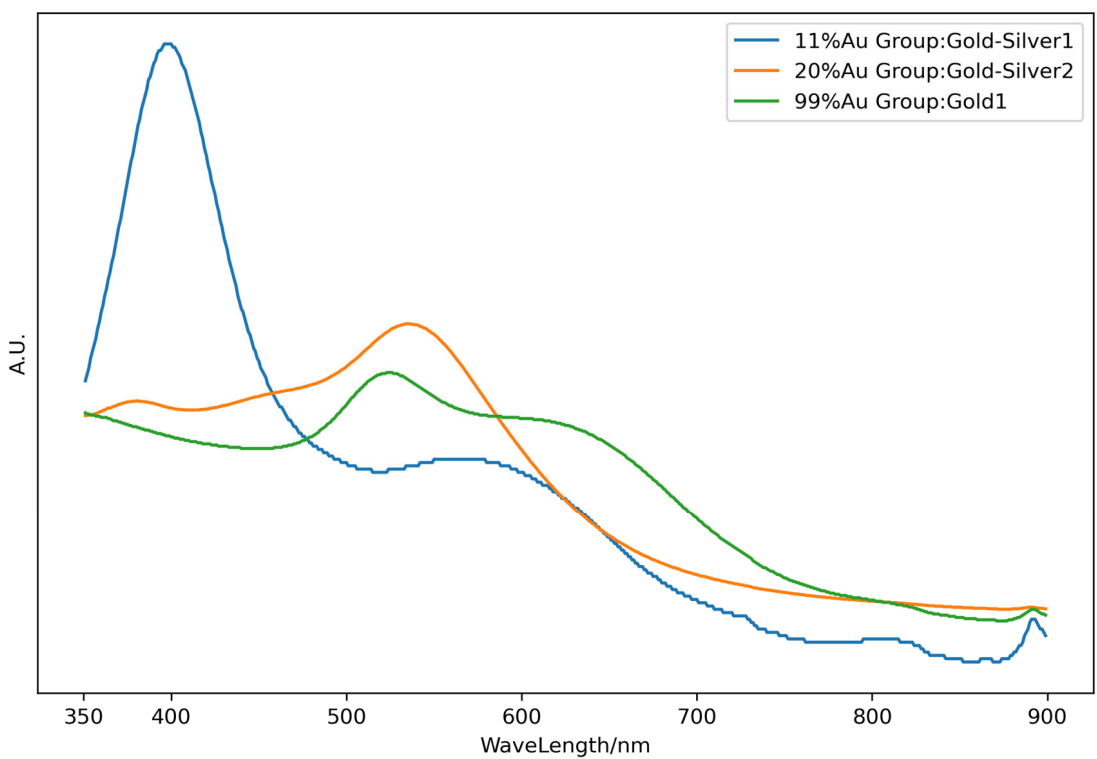

图 3: 三个代表组的紫外可见光光谱（UV-Vis）数据（已归一化）

# 结果与讨论：

  首先我们通过 SEM 对产物直接观察，确定成功合成纳米晶体。并进一步通过 DLS 确定不同组别的粒度统计分布。由于不具备透射电子显微镜（TEM），无法对纳米原子尺度结构进行分析，故使用 UV-Vis 曲线通过对产物光学性质的分析推测其组成及结构。已知下列数据：30.2nm 直径的金纳米晶具有 527nm 的吸收峰[1]；29.1nm 直径的银纳米晶具有 407nm 的吸收峰[1]；约 20nm 直径的含有 53.8%Au 的金银纳米笼具有 760nm 的共振吸收峰[3]。同时，直径越大的纳米晶的吸收峰更加红移[1]；金银纳米笼中，金比例越大吸收峰越红移[3]。

**1，对银离子对金纳米晶合成的影响与其反应机制的讨论：**

  我们选取三组具有代表性的组别进行分析。如图 3 所示，在 Gold-Silver1 组中，金纳米晶的特征峰产生了明显的红移（约 580nm），这可能是由于长时间的加热导致金纳米晶过度生长；其中在约 400nm 位置具有一个非常突出的峰值，对应银纳米晶的特征峰；同时在约 800nm 处有一个可见的峰，与金银纳米笼所对应的吸收波长相接近。考虑到有特征鲜明的新纳米银以及金银纳米笼生成，银离子在合成过程中的反应机制很有可能如图 1 中第二条反应机制所示。

  在 Gold-Silver2 组中，金纳米晶的特征峰在大约 540nm，这对应一个约 50nm 直径大小的金纳米晶[4]。在 380nm 左右有一个对应银纳米晶体的特征峰。同时在金纳米晶与银纳米晶的特征峰中间，具有一个奇特的平坦的平台曲线,这可能是由于新的金银纳米结构所导致。由于这一组别中，已经通过提纯去除大部分还原剂，同时 Ag 离子是通过室温缓慢还原，背后的反应机制很有可能对应图 1 中的第一条反应机制。

  在 Gold 组中，一个明显的峰出现在 630nm 左右，考虑到反应时间有限，这一峰很有可能与低金浓度的金银纳米笼相对应，通过仔细观察，在 800nm 处有一个小的峰值，这亦与高金浓度的金银纳米龙相对应。考虑到反应时间有限，同时具有很有可能的金银纳米笼结构，

这一反应背后的机理很有可能对应图 1 中的第二条反应机制。

图 3 中亦给出了 Au 所占的原子比例，可见 Au 的浓度越大，Ag 所占的特征峰越不明显，同时 Au 比例最大的 Gold 组，在 630nm 处的比例最大，这与金银纳米笼中 Au 比例越高，整体越红移的结论相对应。

从上述三组代表组的制备过程与产物相应性质，我们可以大致得出一些结论：1，当金前体接近反应完全后，加入银前体，银可以与还原剂反应形成比较明显的银纳米晶，同时形成少量金银纳米笼。2，当通过提纯去除体系大部分还原剂后，Ag 离子在室温的缓慢长时间反应下，仍能被还原，形成不明显的 Ag 特征峰，这可能与体系存留的少量还原剂有关；同时这样的缓慢反应有可能促成新的金银纳米结构的产生。3，在金前体与还原剂剧烈反应时，即使加入少量的银前体，银也可能形成银核，被金前体还原生成金银纳米银笼结构。

**2，关于金核银壳结构可能存在的自组装现象及其他聚集现象的讨论：**

从附录中 SEM 图像可以看出，提纯后的金纳米晶的单分散性较好，但是提纯并用超纯水溶解后的金核银壳 1 组仍呈现聚集现象。我们不排除"金银复合纳米晶"存在自组装现象的可能。这一结论有待后续实验验证。

除此之外，通过观察金核银壳 2 实验组，可以发现在丙酮介质中滴加硝酸银在室温放置 12 小时后，聚集为数千纳米粒度的颗粒。由于已经提纯可以排除是前体药物继续反应的情况。对于这个聚集现象，一方面可以归因为丙酮溶剂并非纳米金的良溶剂，另一方面可以归因为硝酸银离子打破了溶液体系的电平衡，使得胶体体系发生聚沉现象。

# 结论:

在本次探索中，我们采用柠檬酸钠无种子法（Turkevich Method）对金银纳米晶进行进一步探索；以银离子对金纳米晶合成过程的影响为出发点，对实验过程中的光谱特性、聚集问题展开分析，得到以下结论:

1. 银离子以不同的浓度与不同的顺序添加对体系产物的结构与特性具有不同的影响，其背后对应着不同的反应机制。
2. 当还原剂充足，反应剧烈时，加入银离子药剂，很有可能产生新的纳米银核；当反应缓慢，还原剂不充足时，银原子可能以特殊的方式参与纳米晶的结构特征，如黏附等方式。
3. 在银离子缓慢参与反应时，连接金、银特征峰的吸收曲线较为平坦，可能存在新的纳米符合结构。
4. 金核银壳结构可能存在特殊的聚集现象，不排除具有自组装机制的可能。

未来进一步研究需要引入透射电镜从原子排列结构的层面进行探索。

# 附录1. DLS图像:

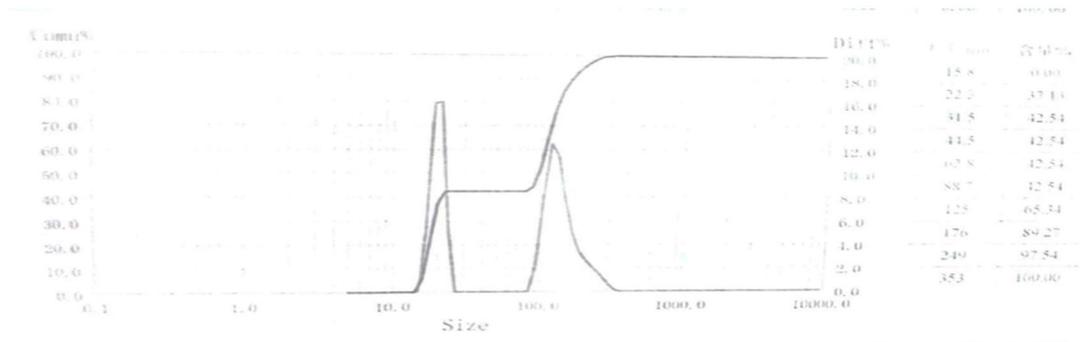

DLS:Gold1

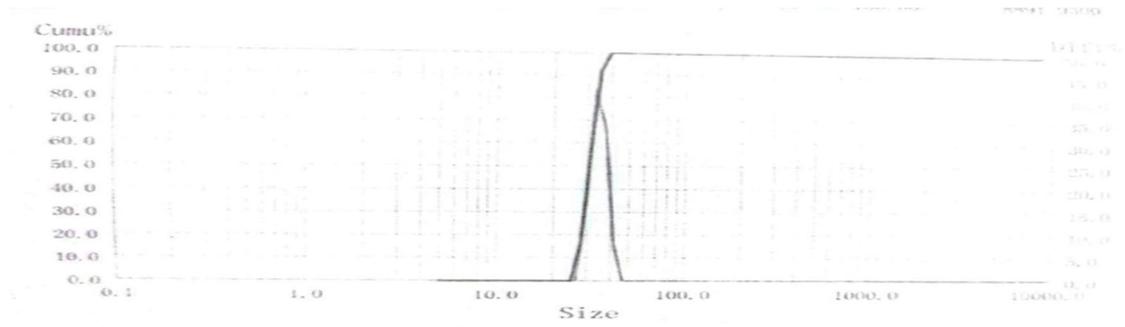

DLS:Gold2

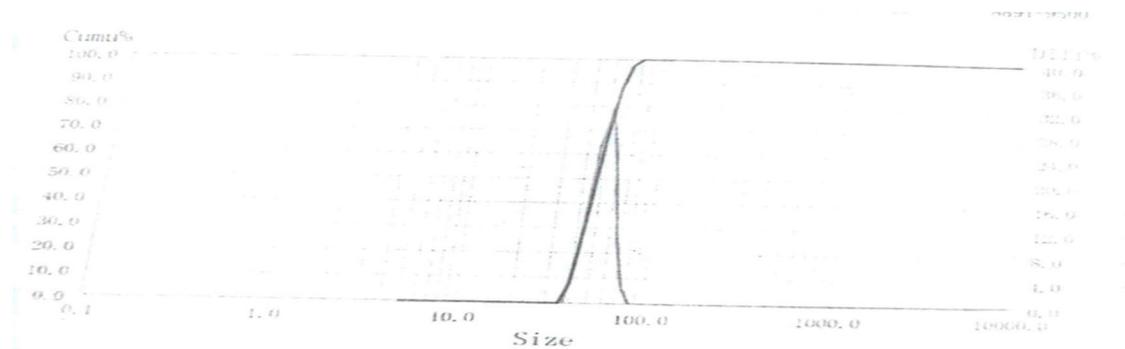

DLS:Silver

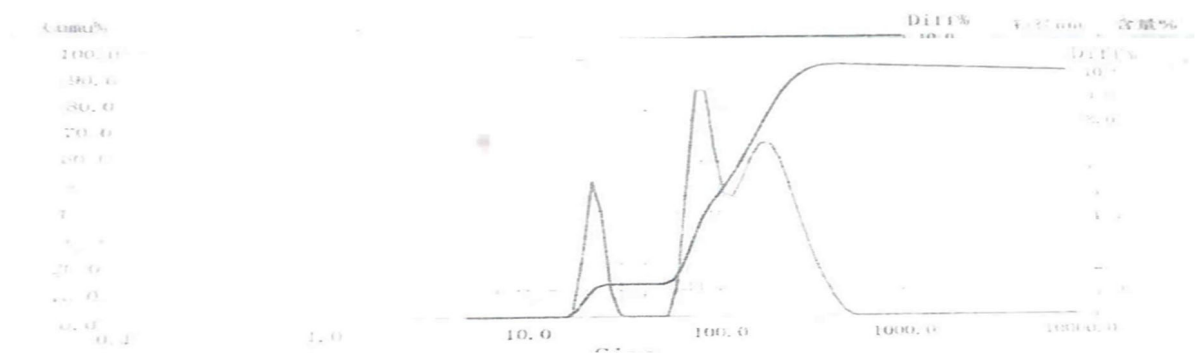

DLS:Gold-Silver1

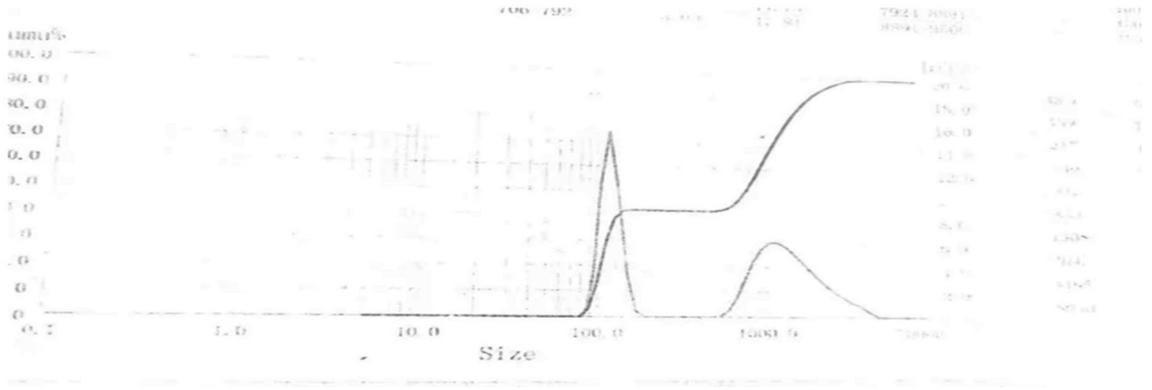

DLS:Gold-Silver2

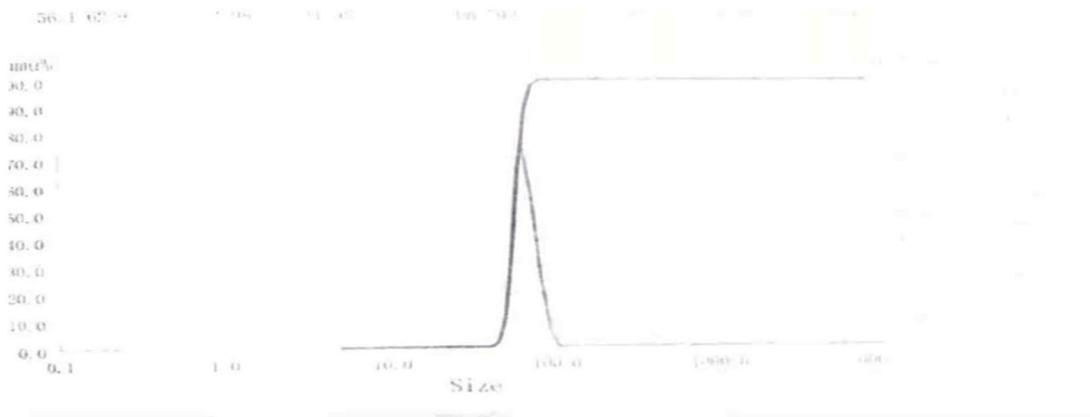

DLS:Silver-Gold1

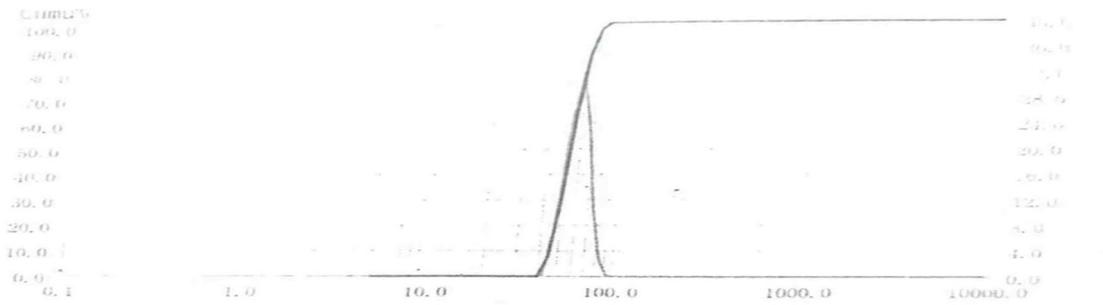

DLS:Silver-Gold2

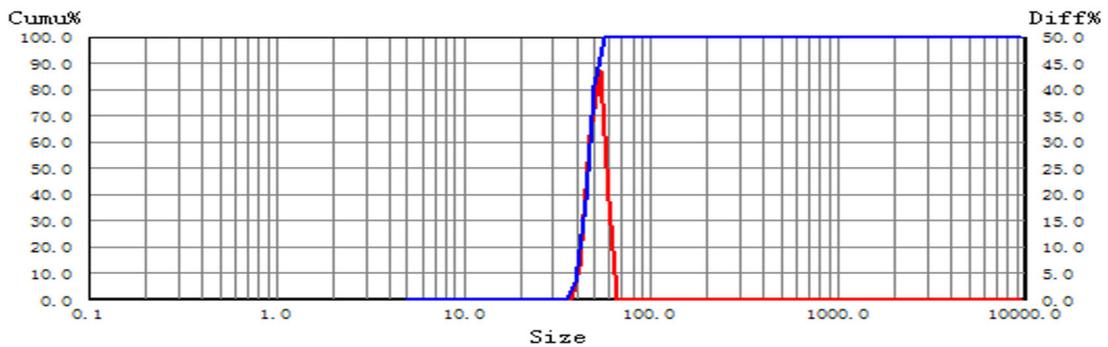

DLS:Silver-Gold3_1

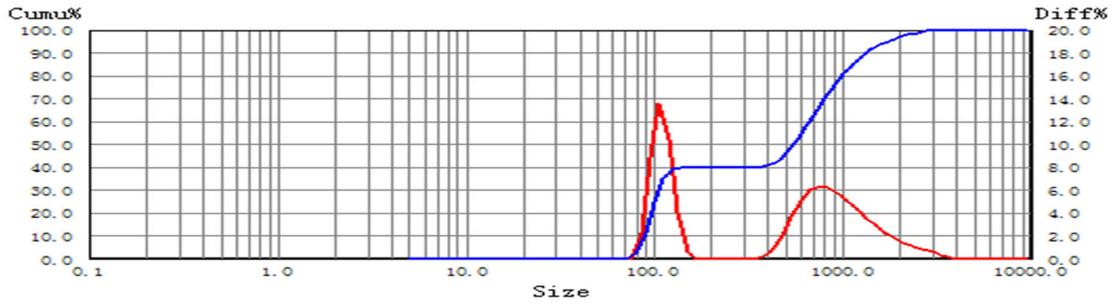

DLS:Silver-Gold3_2

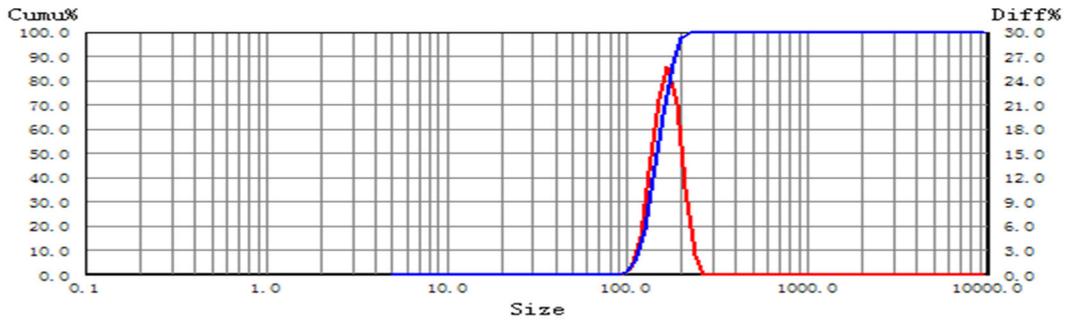

DLS:Silver-Gold3_3 银

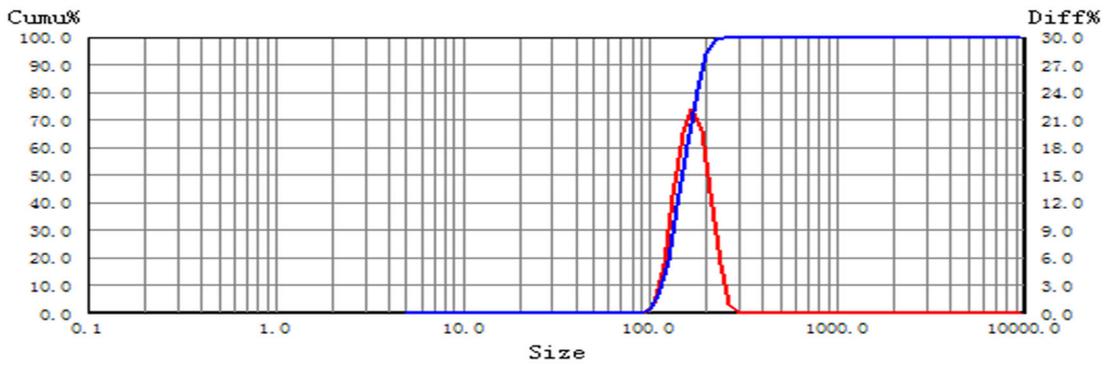

DLS:Silver-Gold3_3 金

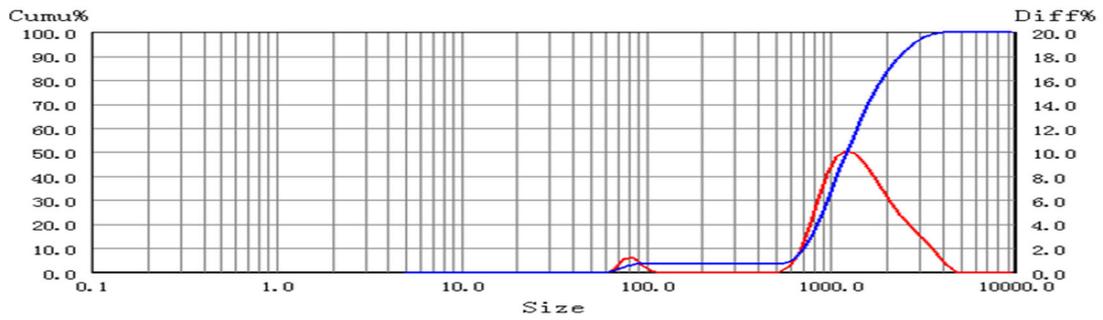

DLS:Silver-Gold3_4 银

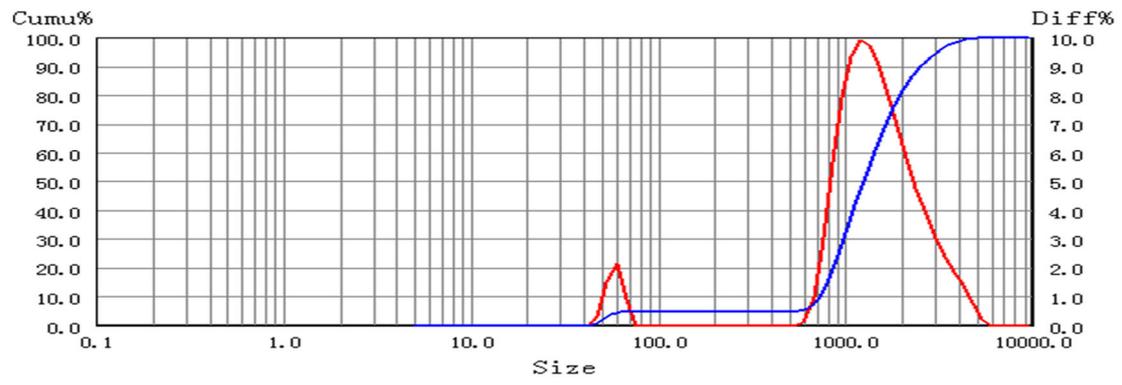

DLS:Silver-Gold3_4 金

# 附录 2. SEM 图像

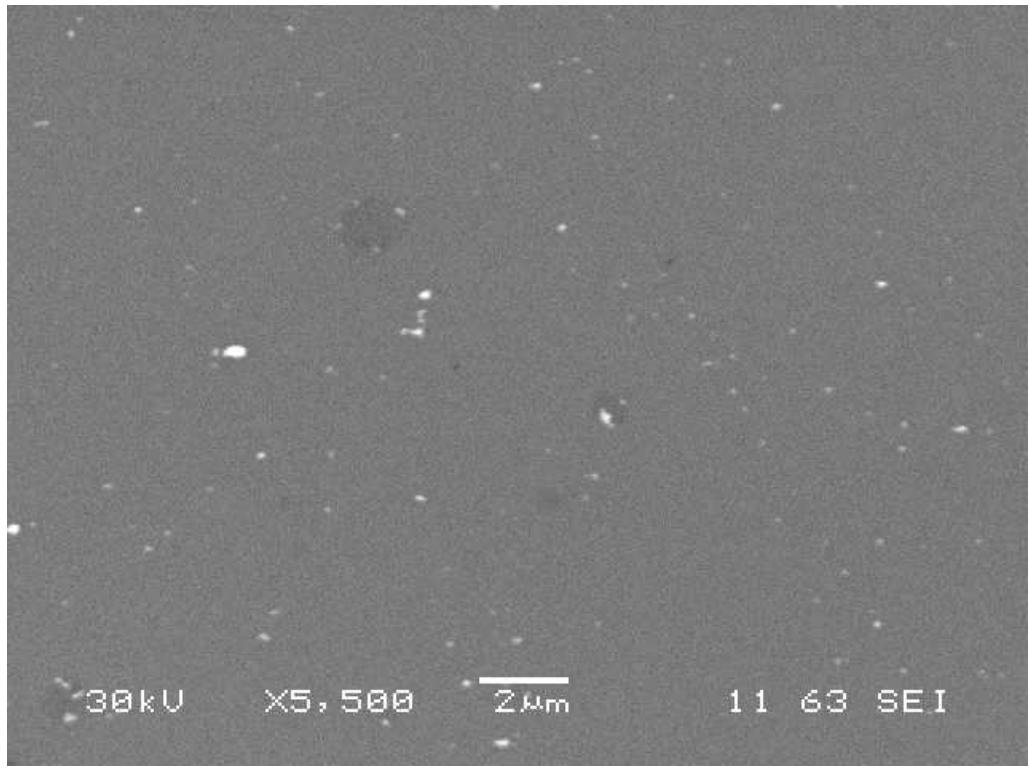

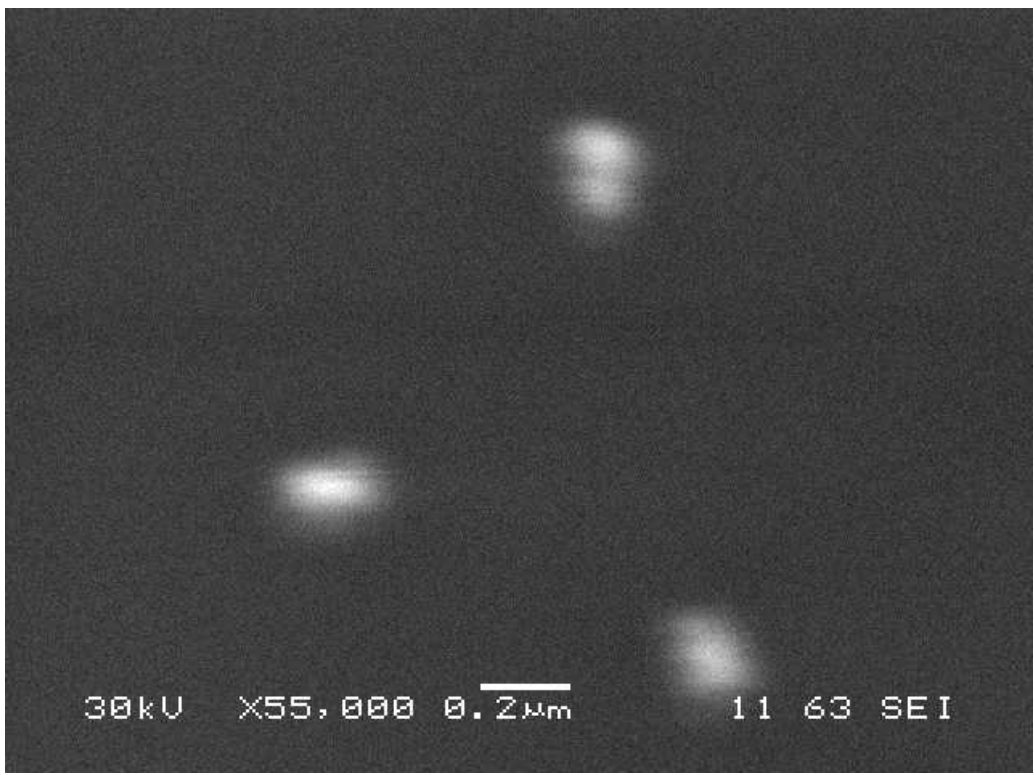

SEM-Gold 组

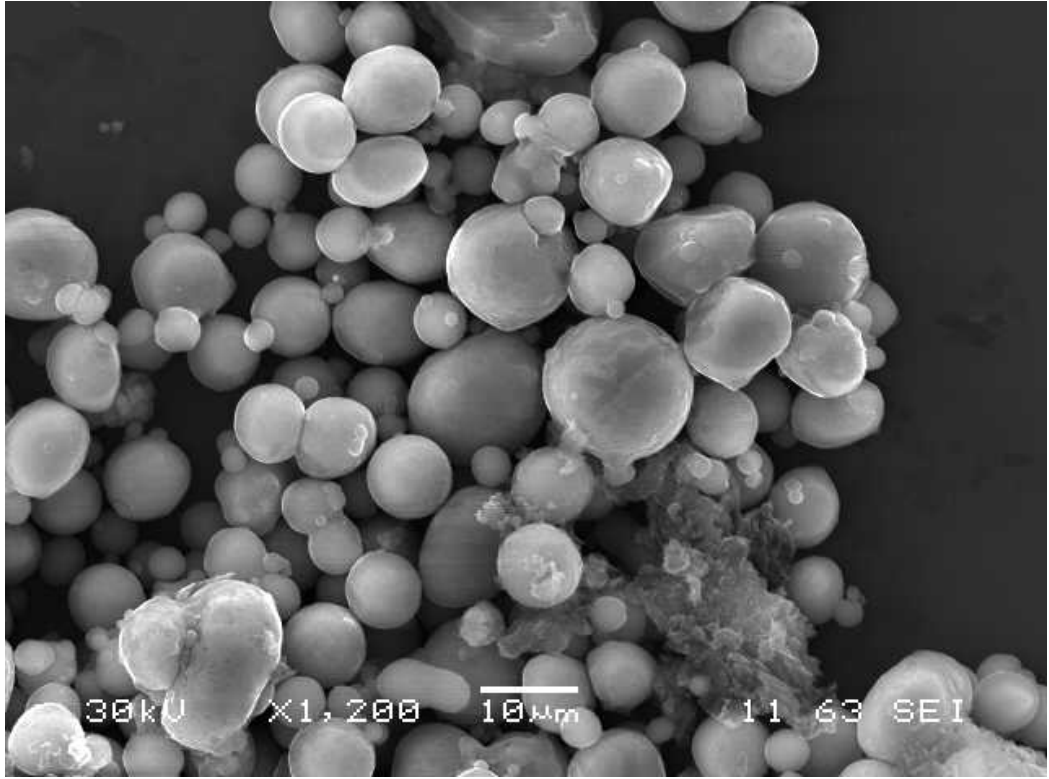

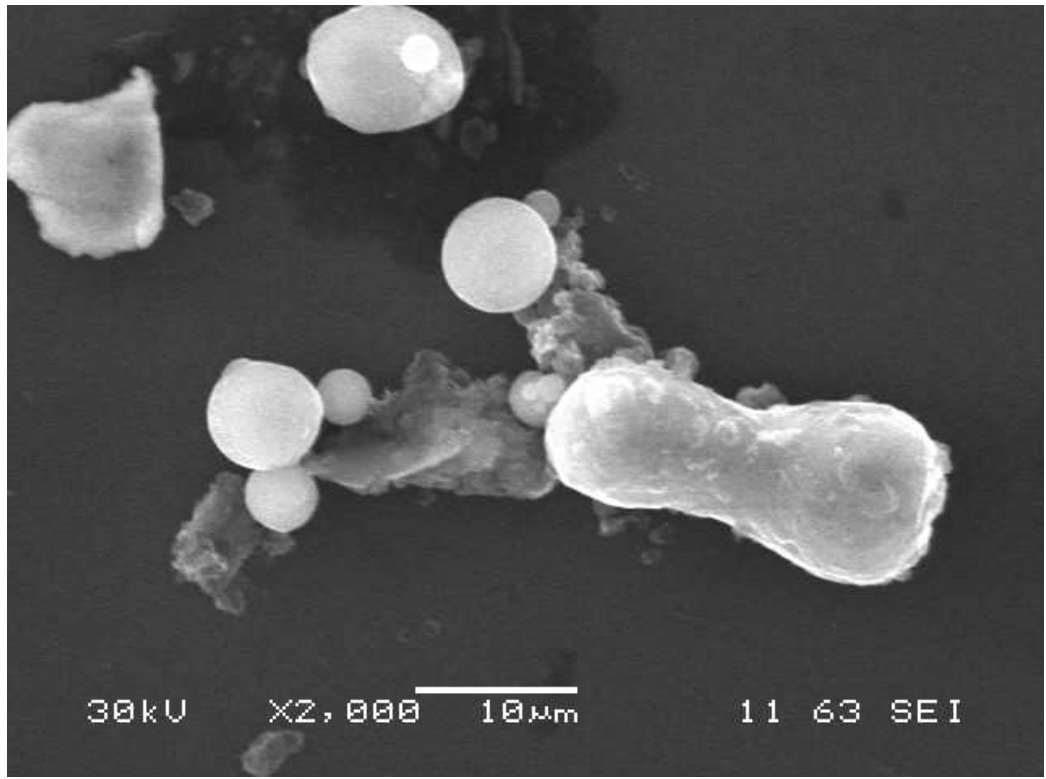

SEM-Gold_2 组

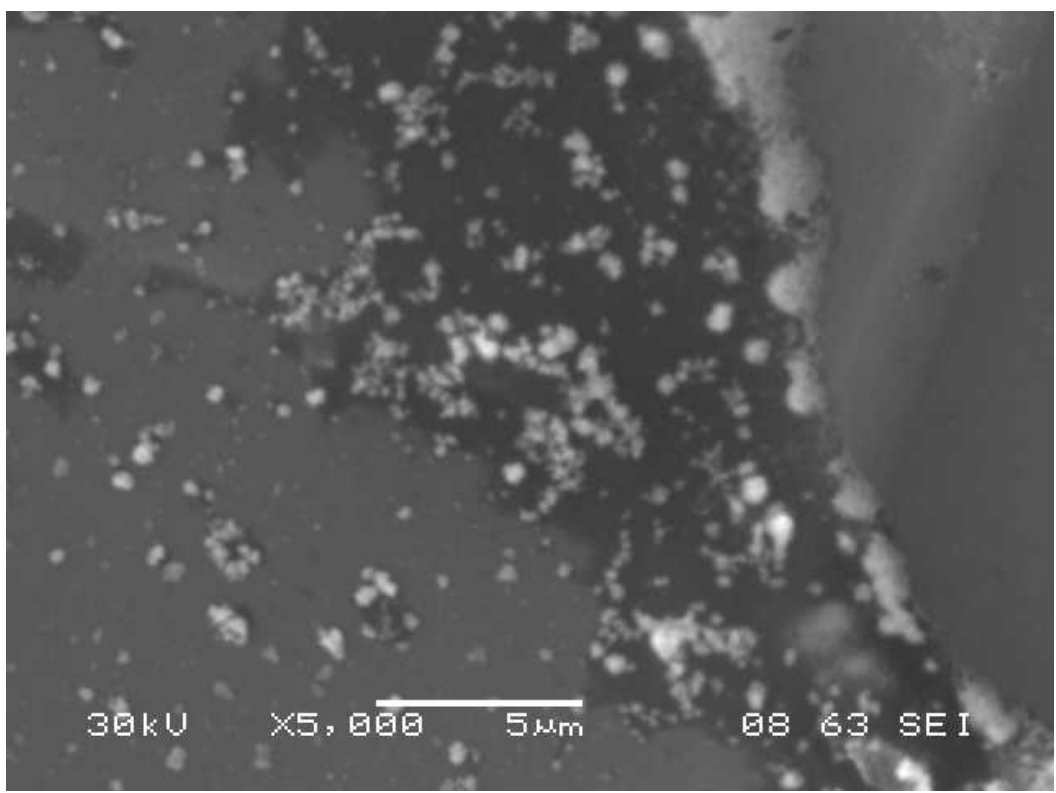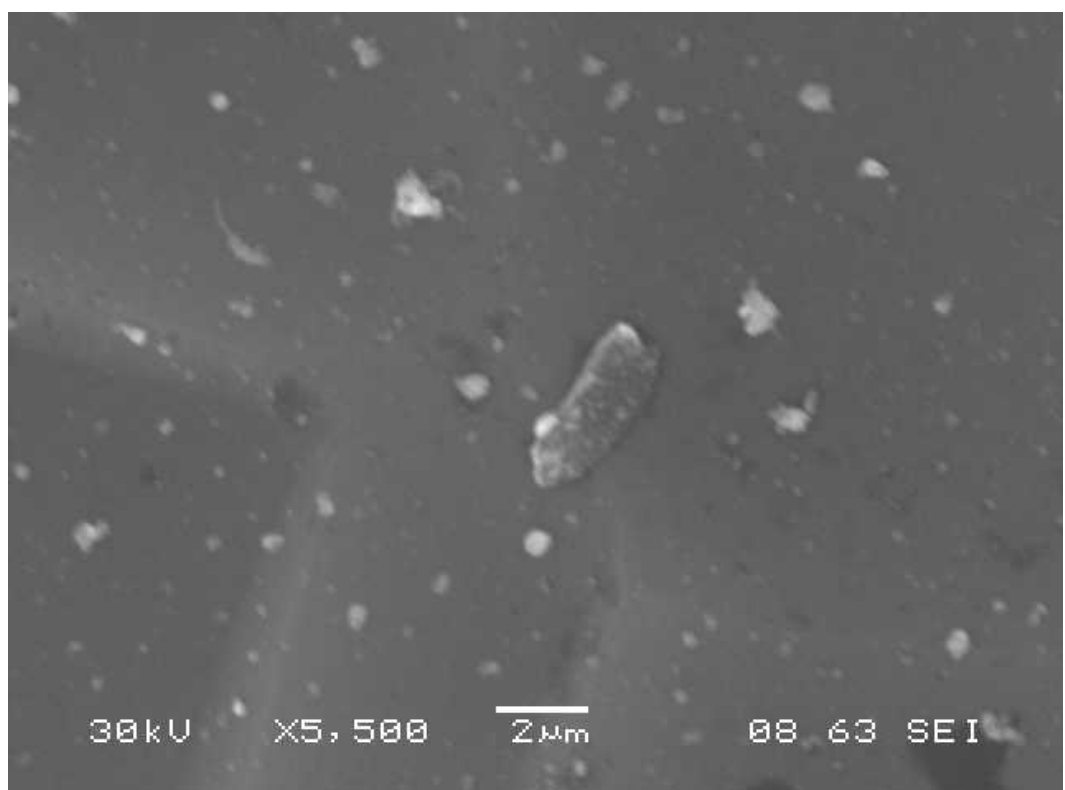

SEM-Gold-Silver 组

# 参考文献：

1. 张继辉, *无种子法合成金纳米晶及核壳结构金银纳米晶的光谱性质研究*. 山东大学, 2014.
2. Turkevich, J., P.C. Stevenson, and J. Hillier, *A study of the nucleation and growth processes in the synthesis of colloidal gold.* Discussions of The Faraday Society, 1951. **11**: p. 55-75.
3. Moreau, L.M., et al., *How Ag nanospheres are transformed into AgAu nanocages.* Journal of the American Chemical Society, 2017. **139**(35): p. 12291-12298.
4. Xia, H., et al., *Synthesis of monodisperse quasi-spherical gold nanoparticles in water via silver (I)-assisted citrate reduction.* Langmuir, 2010. **26**(5): p. 3585-3589.